\documentclass[preprint,3p,twocolumn]{elsarticle}

\usepackage{amssymb}
\usepackage{amsmath}
\usepackage{hyperref}

\journal{Computer Communications}

\begin{document}

\begin{frontmatter}



\title{Digital Signal Processing from Classical Coherent Systems to Continuous-Variable QKD: A Review of Cross-Domain Techniques, Applications, and Challenges} 


\author[1]{Davi Juvêncio Gomes de Sousa\tnoteref{Atext}} 
\ead{davi.juvencio@fieb.org.br}
\author[1,2]{Caroline da Silva Morais Alves} 
\ead{caroline.morais@ufba.br}
\author[1]{Valéria Loureiro da Silva\tnoteref{Atext}} 
\ead{valeria.dasilva@fieb.org.br}
\author[1]{Nelson Alves Ferreira Neto} 
\ead{nelson.neto@fieb.org.br}
\affiliation[1]{organization={QuIIN – Quantum Industrial Innovation EMBRAPII Competence Center CIMATEC in Quantum Technologies},
            organization={SENAI CIMATEC},
            addressline={Av. Orlando Gomes 1845}, 
            city={Salvador},
            postcode={41650-010}, 
            state={Bahia},
            country={Brazil}}
\affiliation[2]{organization={Electrical and Computer Engineering Department, Federal University of Bahia},
            city={Salvador},
            postcode={40210-630},
            state={Bahia},
            country={Brazil}}
\tnotetext[Atext]{Corresponding author.}

\begin{abstract}
This systematic review investigates the application of digital signal processing (DSP) techniques — originally developed for coherent optical communication systems — to continuous-variable quantum key distribution (CV-QKD). The convergence of these domains has enabled significant advances in CV-QKD performance, particularly in phase synchronization, polarization tracking, and excess noise mitigation. To provide a comprehensive and reproducible synthesis of this emerging field, we employed the APISSER methodology, a task-oriented framework adapted from the PRISMA protocol. A structured search across IEEE Xplore and Web of Science databases (2021–2025) yielded 220 relevant publications, which were screened, classified, and analyzed to address six research questions. Our findings highlight that many classical DSP algorithms, such as Kalman filtering, carrier recovery, adaptive equalization, and machine-learning-assisted signal estimation, have been successfully adapted to the quantum regime, often requiring modifications to meet security and noise constraints. We also identify a range of recent DSP innovations in coherent optical communication systems with high potential for future CV-QKD integration, including neural equalization, probabilistic shaping, and joint retiming-equalization filters. Despite these advances, challenges remain in achieving robust phase tracking under ultra-low Signal-to-Noise Ratio (SNR) conditions, real-time polarization compensation, and secure co-existence with classical channels. This review maps current trends, technical barriers, and emerging opportunities at the intersection of signal processing for quantum and classical communication, supporting the development of scalable and resilient CV-QKD systems. 

\end{abstract}



\begin{keyword}
CV-QKD, DSP, Coherent Optical Communication, Systematic Review, APISSER.

\end{keyword}

\end{frontmatter}



\section{Introduction}

The Continuous-Variable Quantum Key Distribution (CV-QKD) systems face inherent limitations imposed by excess noise due to optical channel degradation and detection imperfections, as well as the inherently ultra-low quantum signal levels. To address these challenges, researchers are progressively repurposing Digital Signal Processing (DSP) techniques, initially developed for Coherent Optical Communication (COC) systems, to CV-QKD, tailoring them to enhance system stability, suppress noise, and boost key generation rates, granting resilience to experimental imperfections \cite{Silva2024SBFoton}. 

Relevant works have achieved success by applying COC techniques, such as phase recovery using adaptive filtering algorithms \cite{alsalami2022scalar}, pilotless synchronization methods \cite{matalla2023pilot}, and noise mitigation strategies for discrete modulation formats \cite{pan2024100} to the CV-QKD context, even with low signal intensities, increasing the complexity. Equalization and error correction schemes have also demonstrated effectiveness in long-distance experimental setups \cite{roumestan20226}. More recently, machine learning techniques have also emerged as promising tools for optimizing key extraction and adaptive receiver design \cite{hajomer2022continuous}.

These developments illustrate a broader technological convergence between coherent optical communication and CV-QKD, although a comprehensive understanding of the associated modifications, performance trade-offs, and implementation challenges remains incomplete. In this context, systematic reviews are crucial for synthesizing and critically evaluating the current state of the art in fields characterized by rapid publication growth and diverse technical approaches. Furthermore, by adhering to an explicit and reproducible methodology, such reviews enhance transparency and reliability in identifying research trends, knowledge gaps, and opportunities for future development.

To address these limitations, this study investigates how DSP techniques originating from coherent optical communication systems have been applied to CV-QKD, the adaptations required, and what impacts these techniques have on system performance and practical feasibility. A structured search was conducted in the \textit{IEEE Xplore} and \textit{Web of Science} databases, covering the period from 2021 to 2025. Following the APISSER methodology~\cite{castillo2022apisser} — an approach tailored for systematic reviews in engineering contexts — and applying strict inclusion criteria, this work has selected 220 publications for full-text review and data extraction. The analysis focuses on identifying predominantly adopted DSP strategies and commonly used performance metrics, while documenting reported technical challenges and future research priorities.

The paper is organized as follows: Section~\ref{methodology} details the systematic review methodology, including the APISSER framework and data extraction procedures. Section~\ref{results} presents the results, structured according to the six research questions, highlighting key findings and trends in DSP adaptation for CV-QKD. Section~\ref{conclusions} discusses the main conclusions, summarizing the impact of DSP techniques and outlining future research directions. The acknowledgments and references are provided at the end.

\section{Methodology}
\label{methodology}
To ensure methodological rigor, transparency, and reproducibility, this review adopts the APISSER methodology (\textit{A Priori}, \textit{Plan}, \textit{Identify}, \textit{Screen}, \textit{Select}, \textit{Extract}, \textit{Report}) proposed by Castillo and Grbovic~\cite{castillo2022apisser}. APISSER is a structured, task-oriented framework specifically designed to address the challenges of conducting systematic literature reviews in engineering domains. It is built upon the principles of the PRISMA methodology~\cite{page2021prisma}, yet adapted to the engineering context by incorporating support tools for data-intensive tasks and emphasizing early definition of research scope and data handling procedures.

The methodology is organized into six sequential phases. The initial phases focus on defining the topic, justifying its relevance, and formulating clear research questions. Subsequent steps guide the planning and execution of database searches, selection of eligible publications, and systematic extraction of data. The final phase involves synthesizing and reporting the findings in alignment with the review objectives. In the following subsections, each phase of the APISSER framework is detailed as applied to this review.

\subsection{A priori}

The \textbf{A priori} phase establishes the conceptual foundation of the systematic review. It comprises three essential components: (A1) definition of the research topic, (A2) justification of its relevance, and (A3) formulation of the guiding research questions (RQs).

\textbf{A1 – Topic:}  
This review investigates the application of DSP techniques — originally developed for coherent optical communication systems — to CV-QKD. While DSP plays a central role in modern optical communication systems, especially in mitigating dispersion, phase noise, and other channel impairments, its systematic integration into CV-QKD remains underexplored. CV-QKD, in turn, presents a promising approach to quantum-secure communication but faces substantial challenges related to low signal-to-noise ratios and system instability. The convergence between these domains has produced a growing body of work. However, no comprehensive review exists to assess how DSP has been adapted, evaluated, or optimized in the context of CV-QKD.

\textbf{A2 – Rationale:}  
Despite significant experimental progress, there remains a lack of consolidated knowledge regarding the role of DSP in enhancing CV-QKD systems. Existing studies vary widely in methodology, evaluation metrics, and scope, making it difficult to compare results or assess general trends. A systematic review is therefore necessary to synthesize this fragmented literature, identify recurring challenges, and outline potential research directions. The findings aim to support the development of more robust and scalable CV-QKD systems by leveraging insights from established DSP practices in coherent optical communication.

\textbf{A3 – Research Questions:}  
To address this gap, the following Research Questions were formulated to guide the review:

\begin{itemize}
    \item \textbf{RQ1:} How has the use of DSP techniques from coherent optical communication improved the performance of CV-QKD systems?
    \item \textbf{RQ2:} Which recent DSP innovations in coherent optical communication have not yet been applied to CV-QKD but hold significant potential?
    \item \textbf{RQ3:} What are the future perspectives for the use of DSP in CV-QKD systems?
    \item \textbf{RQ4:} What performance metrics are most commonly used to assess the impact of DSP techniques on CV-QKD?
    \item \textbf{RQ5:} What technical challenges still limit the full integration of DSP into CV-QKD systems?
    \item \textbf{RQ6:} Which DSP techniques from coherent optical communication have been successfully adapted to CV-QKD, and which required specific modifications?
\end{itemize}

\subsection{Plan}

The planning phase defines the operational structure of the systematic review, ensuring its consistency, reproducibility, and methodological rigor. This phase encompasses seven elements, hereafter referred to as P1 through P7, which are detailed below.

\textbf{P1 – Study Characterization:}  
The scope of this review includes publications addressing the application or adaptation of DSP techniques — originally developed for coherent optical communication — to CV-QKD systems. The review focuses on recent advancements from 2021 to 2025, capturing state-of-the-art developments.

The search strategy was based on the following keywords: \textit{“quantum key distribution”, “CV-QKD”, “digital signal processing”, “DSP”, “coherent detection”, “phase recovery”, “modulation”, “equalization”, “machine learning”, “pilotless synchronization”, “signal estimation”}.

Only peer-reviewed journals and conference papers were considered. The selected databases were \textbf{Web of Science} and \textbf{IEEE Xplore}, both widely used in the engineering field. The IEEE Journal of Quantum Electronics and the IEEE Photonics Technology Letters were identified as potential target venues for publication.

\textbf{P2 – Eligibility Criteria:}  
General inclusion and exclusion criteria are shown in Table~\ref{tab:general_criteria}. In addition, specific inclusion criteria were defined for each research question (RQ) to allow for a segmented and objective assessment of the selected publications.

\begin{table}[ht]
\centering
\caption{General inclusion and exclusion criteria}
\label{tab:general_criteria}
\small
\begin{tabular}{|c|c|p{4cm}|}
\hline
\textbf{Type} & \textbf{Code} & \textbf{Description} \\ \hline
Inclusion & INC-G1 & The article addresses CV-QKD systems. \\ \hline
Inclusion & INC-G2 & The article discusses DSP techniques. \\ \hline
Inclusion & INC-G3 & The article presents analysis, a technical proposal, or application. \\ \hline
Inclusion & INC-G4 & Published between 2021 and 2025. \\ \hline
Inclusion & INC-G5 & Available in full text. \\ \hline
Exclusion & EXC-G1 & Focuses exclusively on DV-QKD. \\ \hline
Exclusion & EXC-G2 & Does not present relevant technical content. \\ \hline
Exclusion & EXC-G3 & Full text is not accessible. \\ \hline
Exclusion & EXC-G4 & Duplicate of an already included article. \\ \hline
Exclusion & EXC-G5 & Irrelevant to the scope despite containing related terms. \\ \hline
\end{tabular}
\end{table}

\textbf{P3 – Data Items:}  
For each selected publication, a set of data items was extracted and linked to the corresponding research questions. These items include: (i) type of DSP technique applied, (ii) system architecture, (iii) performance metrics used, (iv) experimental setup or simulation parameters, (v) observed limitations, and (vi) proposed future developments. Each data item was catalogued in the local database (L-DB) for structured analysis.

\textbf{P4 – Database Search Strategy:}  
Search queries were executed manually in both IEEE Xplore and Web of Science. Logical combinations of the selected keywords were used with filters for publication date (2021–2025) and document type (conference/journal articles). Search results were exported in machine-readable CSV format and imported into the local database.

\textbf{P5 – Selection Process:}  
The selection process was conducted in three stages: (i) title and abstract screening, (ii) eligibility check based on full text, and (iii) final inclusion based on relevance to RQs. This procedure is detailed in Section~\ref{ScreenAndSelect} and visualized in the PRISMA flow diagram.

\textbf{P6 – Tools and Data Management:}  
The Local Database was implemented in \texttt{SQLite} and fully managed through custom \texttt{Python} scripts. The database schema was designed to store comprehensive publication metadata, eligibility decisions, extracted data items, and explicit mappings to the defined Research Questions (RQs). Automated routines were developed to identify and merge duplicate entries based on DOI and normalized title-year keys. All scripts, datasets, and configuration files were maintained under \texttt{Git} version control to guarantee transparency, reproducibility, and full traceability of the review process. The complete toolkit supporting this phase is openly available at:  
\url{https://github.com/CCTQ-CIMATEC/cvqkd-dsp-review-tools}

\textbf{P7 – Systematic Review Protocol:}  
All planning details — including eligibility criteria, keyword strategy, data extraction templates, and database schema — were compiled in an internal protocol document. This protocol was reviewed and validated by the research team before execution.

\subsection{Identify}

In this phase, the systematic search and initial organization of the publication dataset were performed. The process consisted of three steps: database search queries, data export, and the construction of an L-DB for structured processing.

\textbf{I1 – Database Search Queries:}  
Searches were conducted in the \textit{IEEE Xplore} and \textit{Web of Science} databases using logical combinations of descriptors related to CV-QKD, DSP, and Coherent Optical Communication. The search strings were tailored to the syntax of each platform and designed to reflect the focus of each research question (RQ), ensuring alignment with the inclusion criteria (INC-G1 to INC-G5). Table~\ref{tab:search_strings} summarizes the search expressions applied.

\begin{table}[ht]
\centering
\caption{Search strings by research question (RQ)}
\label{tab:search_strings}
\footnotesize
\begin{tabular}{|c|p{6.2cm}|}
\hline
\textbf{RQ} & \textbf{Search Expression} \\ \hline
RQ1 & ("CV-QKD" OR "continuous-variable quantum key distribution") AND ("DSP" OR "digital signal processing") \\ \hline
RQ2 & "coherent optical communication" AND "digital signal processing" AND (innovation OR novel OR "emerging techniques") \\ \hline
RQ3 & ("CV-QKD" OR "continuous-variable quantum key distribution") AND "digital signal processing" AND (innovation OR evolution OR "emerging techniques" OR "next-generation" OR future OR perspectives OR trends) \\ \hline
RQ4 & ("CV-QKD" OR "continuous-variable quantum key distribution") AND "digital signal processing" AND ("performance evaluation" OR "key rate" OR SNR OR "bit error rate") \\ \hline
RQ5 & ("CV-QKD" OR "continuous-variable quantum key distribution") AND "digital signal processing" AND (limitation OR noise OR impairment OR nonlinearity OR complexity OR "trade-off") \\ \hline
RQ6 & ("CV-QKD" OR "continuous-variable quantum key distribution") AND ("carrier phase estimation" OR "homodyne detection" OR "coherent receiver" OR equalization OR "pilot tone" OR "clock recovery") \\ \hline
\end{tabular}
\end{table}

\textbf{I2 – Export Publication Data:}  
Search results were exported in machine-readable CSV format. For each query, the metadata of all retrieved publications — including title, authors, abstract, DOI, keywords, publication year, and source — were saved. These structured datasets enabled automated filtering, deduplication, and integration into the local database.

\textbf{I3 – Local Database Construction:}  
All exported CSV files were parsed using Python scripts and imported into a structured local database (SQLite format). Each search result was loaded into a separate table to maintain traceability. The L-DB schema was extended with fields for eligibility status, RQ association, duplication flags, and extracted data items. Duplicate records were identified and marked using automated string-matching algorithms, ensuring that each unique publication was reviewed only once. 

To guarantee reproducibility and facilitate auditability, the entire process — including parsing scripts, database schema, and intermediate datasets — was version-controlled using Git. This setup supported consistency across screening and extraction phases and allowed collaborative editing with full traceability of changes.

\subsection{Screen}
\label{ScreenAndSelect}

The screening process was conducted in two sequential stages, following the APISSER guidelines. In the \textbf{basic screening} stage (S1), each publication was evaluated based on its title, authorship, and abstract. This initial review determined whether the general inclusion criteria (INC-G1 to INC-G5) were potentially met. Notes were added to the L-DB under the screen basics notes (\texttt{SBN}) field to document partial observations and uncertainties.

In the subsequent \textbf{overview screening} stage (S2), the remaining inclusion criteria were assessed through a general review of the full text. Articles that passed the basic screening were downloaded, and reviewers conducted a high-level evaluation of their content without performing full data extraction. At this stage, the fields screen overview notes (\texttt{SON}) and all remaining inclusion flags in the L-DB were completed. The goal was to determine whether each article provided sufficient alignment with the RQs, technical relevance, and DSP application context.

\subsection{Select}

In the final selection stage (S3), studies that satisfied all inclusion criteria were marked as selected (\texttt{SEL} = true) in the L-DB and proceeded to full-text analysis and data extraction. Duplicate and irrelevant entries were identified and removed using automated scripts during the import phase, with manual confirmation performed in the structured review spreadsheet during the screening stages. The selection process was thus supported by a combination of the screening spreadsheet and the Python-based command-line toolkit, which enabled systematic handling of inclusion/exclusion criteria and export of the resulting datasets for further analysis.

\begin{figure}[hbt]
\centering
\includegraphics[width=0.50\textwidth]{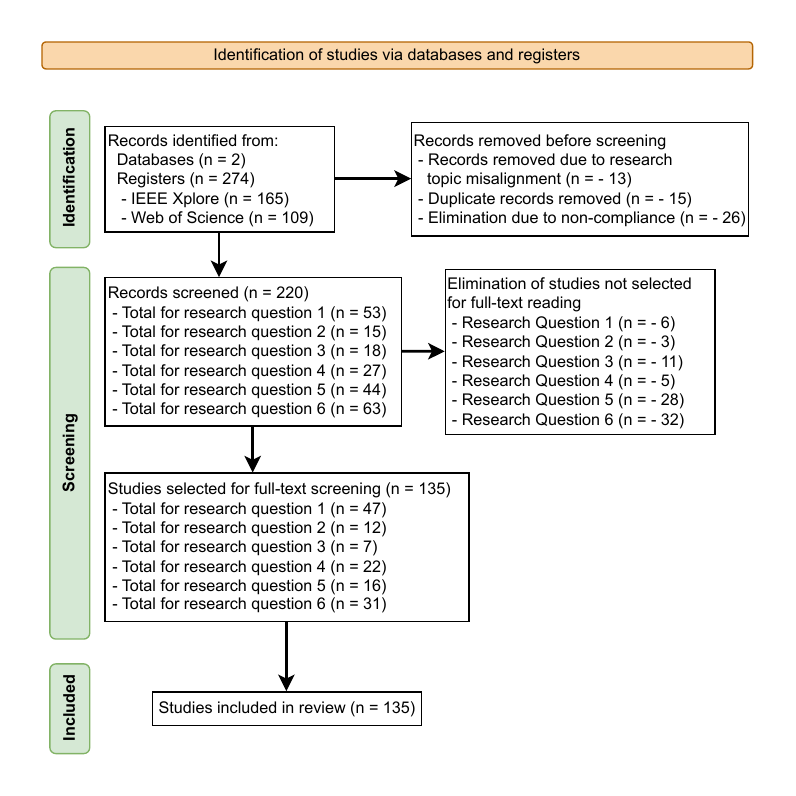}
\caption{PRISMA flow diagram for the screening and selection process.}
\label{fig:prisma}
\end{figure}

Figure~\ref{fig:prisma} provides a detailed overview of the identification, screening, and inclusion process, following the PRISMA 2020 guidelines. A total of 274 records were initially identified across two major databases: IEEE Xplore (165 records) and Web of Science (109 records). Before screening, 54 records were excluded due to various issues: 13 for topic misalignment, 15 due to duplication, and 26 for non-compliance with general inclusion criteria.

Following this initial filtering, 220 studies proceeded to the screening phase. These were mapped across the six research questions, with varying relevance: RQ1 (53 studies), RQ2 (15), RQ3 (18), RQ4 (27), RQ5 (44), and RQ6 (63). During the screening process, 85 studies were excluded from full-text analysis due to insufficient alignment with the RQs or technical depth. The breakdown of these exclusions is shown in the diagram, with RQ5 and RQ6 exhibiting the highest rates of preliminary rejection (28 and 32 studies, respectively).

Ultimately, 135 studies were selected for full-text screening and were retained for inclusion in the review. These included: RQ1 (47 studies), RQ2 (12), RQ3 (7), RQ4 (22), RQ5 (16), and RQ6 (31). The final inclusion count reflects the rigorous application of eligibility criteria across all phases of the review, ensuring that only high-quality and relevant sources were analyzed in depth.

\subsection{Extract}

Following the selection phase, full-text reading and data extraction were conducted for each of the selected publications. This process corresponds to the extraction phase (E1) of the APISSER methodology and aims to populate the predefined data fields in the L-DB.

All items were extracted manually through a comprehensive reading of each article. Access to the full texts was facilitated via DOI links, and articles were downloaded and stored in a local repository. The extracted data were inserted into the L-DB using a custom Python graphical interface (\texttt{ex\_gui.py}), which displayed basic metadata and allowed for structured input of data fields and category labels.

To support subsequent analysis and facilitate comparative synthesis, each article was assigned a classification category recorded under the \texttt{SLC} (Selection Category) field. The following categories were defined:

\begin{itemize}
    \item \textbf{C1:} Fully described DSP implementation in CV-QKD
    \item \textbf{C2:} Referenced or partially described DSP method
    \item \textbf{C3:} Review paper or theoretical proposal
\end{itemize}

Table~\ref{tab:itens_dados} lists the data items defined for extraction, including the mapping to specific research questions. Each field was recorded as a structured text or numerical entry in the L-DB. These items supported both quantitative and qualitative analyses in the results and discussion sections of this review.

\begin{table}[ht]
\centering
\caption{Data items extracted from the studies}
\label{tab:itens_dados}
\small
\begin{tabular}{|c|p{4cm}|c|}
\hline
\textbf{Code} & \textbf{Data Item} & \textbf{RQs} \\ \hline
D1 & Type of DSP technique applied & RQ1, RQ6 \\ \hline
D2 & Purpose of the technique (e.g., phase correction, equalization) & RQ1, RQ5 \\ \hline
D3 & Evaluation metrics used & RQ4 \\ \hline
D4 & Quantitative results & RQ1, RQ4 \\ \hline
D5 & Technical challenges discussed & RQ5 \\ \hline
D6 & Innovations in DSP for Coherent Optical Communication with potential applications in CV-QKD & RQ2 \\ \hline
D7 & Potential future applications mentioned & RQ3 \\ \hline
D8 & Adaptation of Coherent Optical Communication techniques to CV-QKD & RQ6 \\ \hline
\end{tabular}
\end{table}

\section{Results}
\label{results}
This section presents the results obtained through the application of the systematic protocol described in the methodology, based on the six previously defined RQs. The analysis was conducted on a set of 220 publications selected between 2021 and 2025 from the IEEE Xplore and Web of Science databases.

\subsection{RQ1 – How has the use of DSP techniques from Coherent Optical Communication improved the performance of CV-QKD systems?}

In \cite{schiavon2023high}, a high-rate CV-QKD system is presented in which the digital signal processing (DSP) chain is significantly enhanced by pilot-assisted techniques originally developed for coherent classical communication systems. In particular, the adoption of \textit{root-raised cosine} (RRC) pulse shaping plays a crucial role in mitigating inter-symbol interference (ISI) under bandwidth-limited conditions, while the combination of \textit{optical single-sideband} (OSSB) modulation with RF-heterodyne detection enables the simultaneous recovery of both quadratures using a single balanced detector pair.  

\begin{figure}[hbt]
\centering
\includegraphics[width=0.50\textwidth]{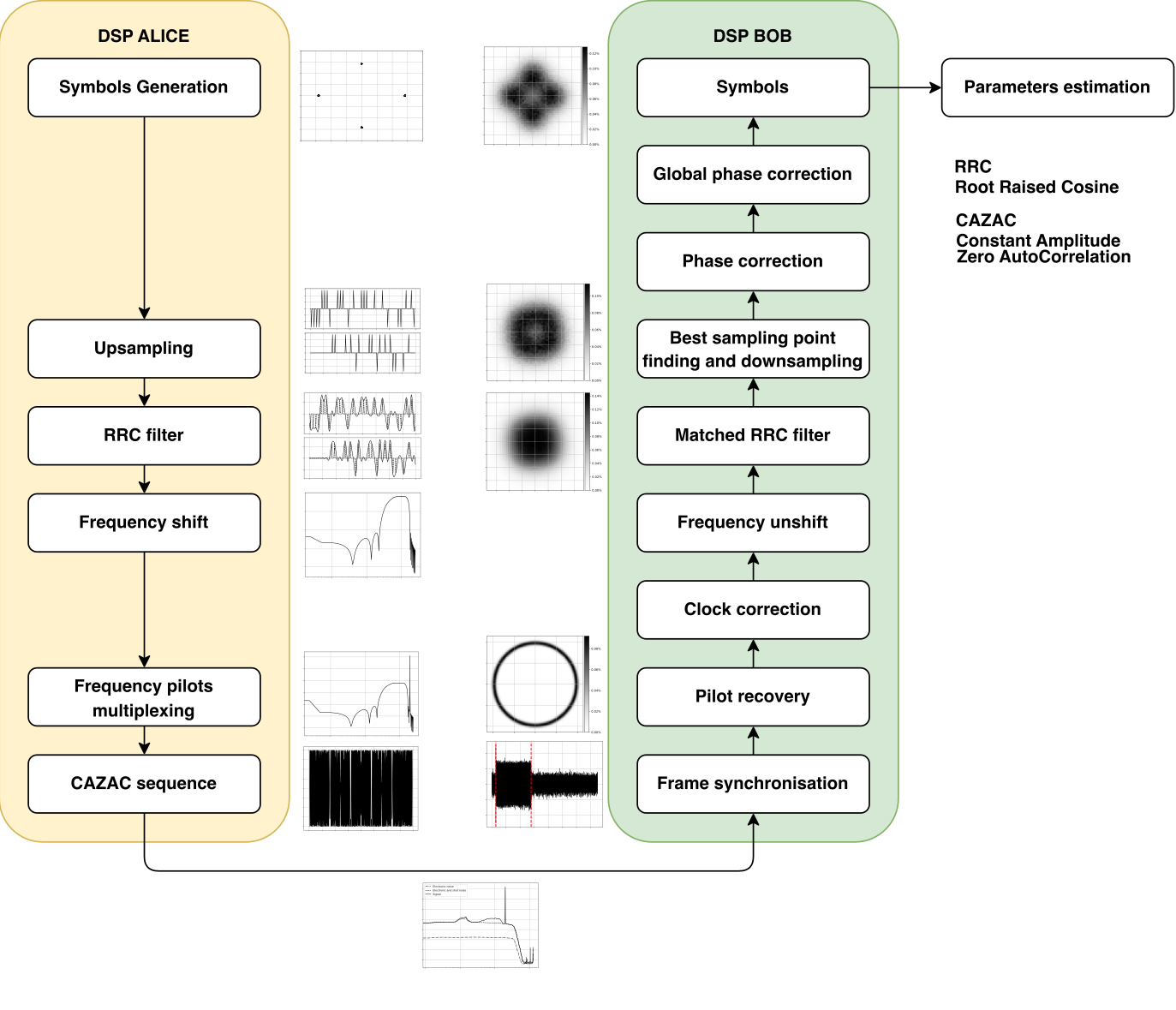}
\caption{Block diagram of the digital signal processing (DSP) used in the CV-QKD system, showing synchronization, pilot-based frequency and phase recovery, matched filtering, and resampling. Reproduced from \cite{schiavon2023high}.}
\label{fig:dsp_scheme}
\end{figure}

The DSP block diagram employed in the system is shown in Fig.~\ref{fig:dsp_scheme}, reproduced from \cite{schiavon2023high}. A clear separation can be observed between the modules implemented at Alice and at Bob. On the transmitter side, this includes symbol generation and filtering, frequency-multiplexed pilot insertion, and the addition of a CAZAC/Zadoff-Chu synchronization sequence. On the receiver side, the DSP pipeline encompasses frame synchronization, pilot-based frequency and phase recovery, matched RRC filtering, and adaptive resampling.  

Through careful optimization of DSP parameters, such as pilot amplitude and RRC roll-off factor, the authors report a performance of 0.15 secure bits per symbol in a back-to-back configuration, corresponding to a key rate of approximately 15~Mbit/s. Furthermore, a proof-of-concept experiment employing a photonic integrated circuit (PIC) receiver demonstrated excess noise levels as low as 0.1~SNU over a 7~km equivalent fiber channel, enabling secure key rates of 280~kbit/s in the asymptotic regime and 123~kbit/s in the finite-size regime. These results underscore the effectiveness of advanced DSP in enhancing both system stability and achievable key rates in practical CV-QKD implementations.

\begin{figure}[hbt]
\centering
\includegraphics[width=0.40\textwidth]{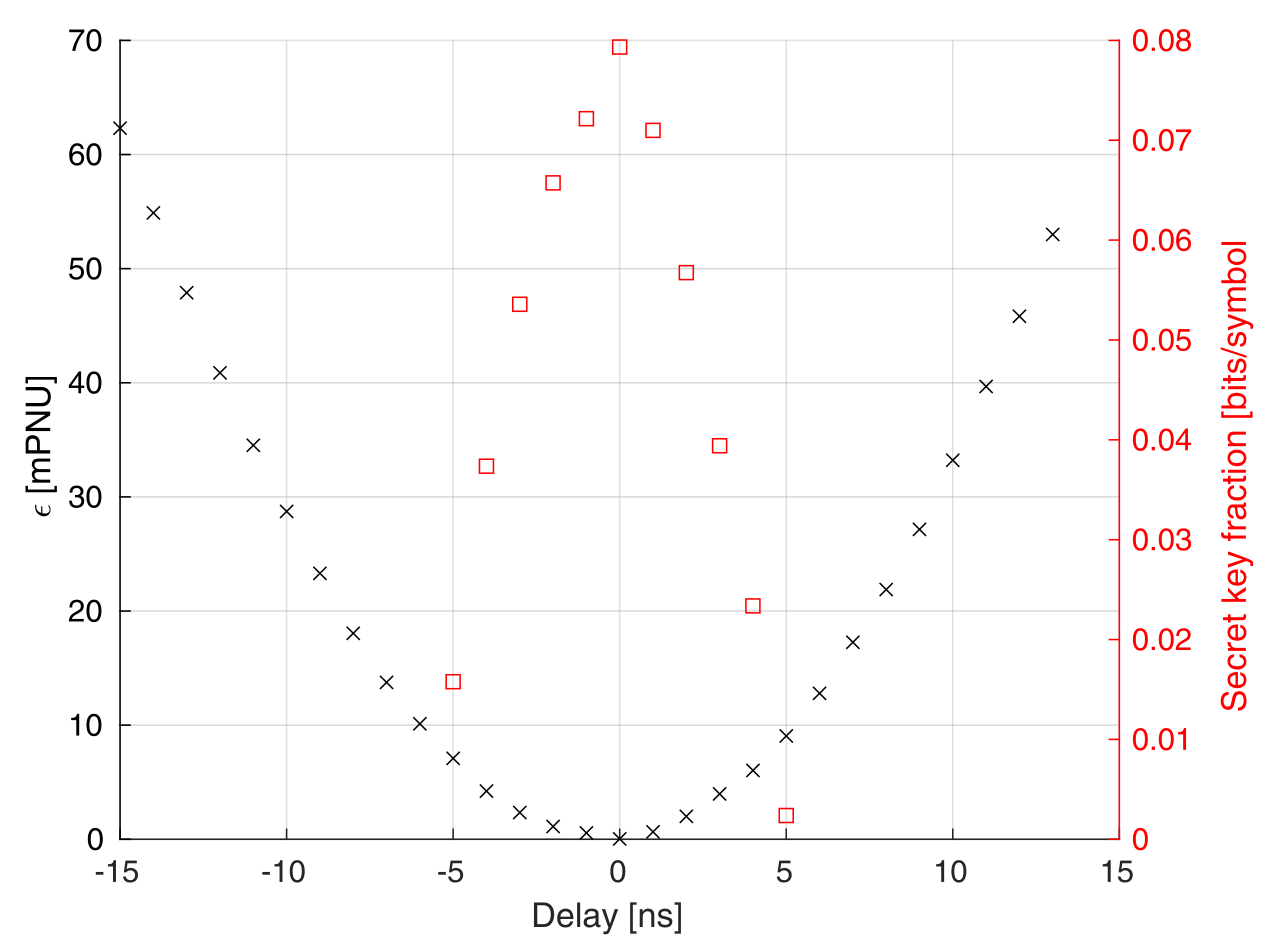}
\caption{Impact of forced erroneous synchronization on excess noise and asymptotic secret key fraction in a clock-synchronous CV-QKD system transmitting over 20~km fiber. Even small deviations from the correct delay can prevent key generation. Reproduced from \cite{chin2022digital}.}
\label{fig:excess_noise_vs_delay}
\end{figure}

In \cite{chin2022digital}, the authors propose and experimentally validate a comprehensive digital synchronization scheme tailored for CV-QKD systems employing a locally generated local oscillator. The method adapts several classical DSP techniques from coherent optical communications to the CV-QKD domain, including timing error detection and clock recovery with auxiliary high-power QPSK signals, frequency offset estimation using dual pilot tones, and phase noise compensation based on Unscented Kalman Filtering (UKF) combined with the M-th power algorithm. Additionally, frame synchronization is achieved through cross-correlation with CAZAC sequences multiplexed into the quantum signal. The authors demonstrate that this approach enables reliable synchronization even in free-running configurations without shared clocks or hardware triggers. The criticality of precise synchronization is exemplified in Fig.~\ref{fig:excess_noise_vs_delay}, which shows that a synchronization delay deviating by more than approximately 1/10 of the symbol duration can prevent any secret key generation, due to an increase in excess noise. Experimental results over 10 km and 20 km of standard single-mode fiber show that the proposed DSP chain yields excess noise levels comparable to externally synchronized systems and supports secret key generation with no performance degradation, achieving secret key fractions of up to 0.074 bits/symbol. The scheme is modulation-format agnostic and requires no additional hardware, making it a promising solution for scalable, cost-effective CV-QKD deployments.

In \cite{pan2023simple}, the authors introduce a Simple and Fast Polarization Tracking Algorithm specifically designed for LLO-based CV-QKD systems. The approach leverages classical coherent optical communication techniques by adopting an orthogonal pilot-tone scheme for polarization estimation and compensation. One pilot tone is employed to estimate the polarization rotation angle, while the second is used to track the phase noise induced by polarization perturbations.  

The optical fiber channel is modeled through a unitary Jones matrix $J(t)$, which captures the time-varying polarization rotation and associated phase fluctuations:  

\begin{equation}
J(t) =
\begin{bmatrix}
\cos \alpha(t) e^{j\varphi_1(t)} & -\sin \alpha(t) e^{j\varphi_2(t)} \\
\sin \alpha(t) e^{-j\varphi_2(t)} & \cos \alpha(t) e^{-j\varphi_1(t)}
\end{bmatrix},
\end{equation}

where $\alpha(t)$ denotes the polarization rotation angle, and $\varphi_{1,2}(t)$ represent phase terms arising from stochastic birefringence in the transmission link~\cite{pan2023simple}.  

Instead of resorting to iterative optimization procedures such as the Constant Modulus Algorithm (CMA), the proposed method exploits the pilot-tone measurements to directly compute the inverse Jones matrix $J^{-1}(t)$. The demultiplexing operation is thus performed according to

\begin{equation}
\begin{bmatrix}
E_V(t) \\
E_H(t)
\end{bmatrix}
=
J^{-1}(t)
\begin{bmatrix}
E^V_{Q,PT1,PT2}(t) \\
E^H_{Q,PT1,PT2}(t)
\end{bmatrix},
\end{equation}

allowing efficient separation of the quantum signal from the pilot tones~\cite{pan2023simple}.  

To mitigate residual polarization-induced impairments due to imperfect estimation of $\alpha(t)$ and $\Delta\varphi(t)$, the scheme incorporates a real-valued finite impulse response (FIR) filter, adaptively trained using the Least Mean Squares (LMS) algorithm. This DSP stage effectively suppresses residual cross-polarization noise, thereby reducing excess noise in the recovered quadratures.  

\begin{figure}[hbt]
\centering
\includegraphics[width=0.40\textwidth]{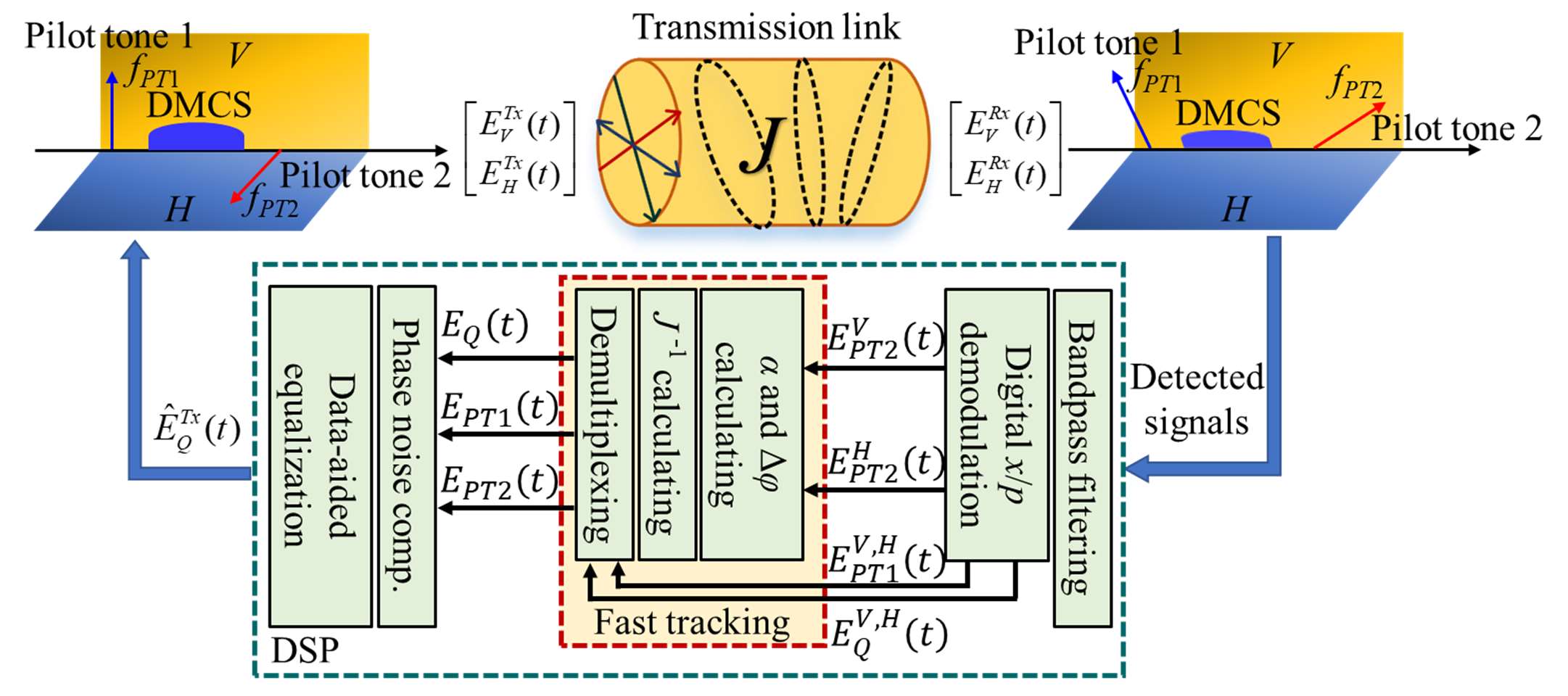}
\caption{Orthogonal pilot-tone-assisted fast polarization tracking algorithm schematic. Reproduced from \cite{pan2023simple}.}
\label{fig:orthogonal_pilot_assisted_polarization_tracking}
\end{figure}

The overall algorithmic workflow, illustrated in Fig.~\ref{fig:orthogonal_pilot_assisted_polarization_tracking}, integrates bandpass filtering, quadrature demodulation, polarization demultiplexing via $J^{-1}$, phase-noise compensation, and adaptive equalization. Experimental validation with a 1-GBaud DG-256QAM CV-QKD system over 24.49 km of standard single-mode fiber confirmed stable performance under polarization scrambling rates up to $12.57~\text{krad/s}$, while numerical simulations extended the tracking capability to $188.50~\text{Mrad/s}$. The results demonstrate lower excess noise and higher secret key rates relative to CMA and FIR-based baselines, establishing the method’s practical suitability for high-speed CV-QKD deployment.

In \cite{chin2024cvqkd}, the authors successfully demonstrate a polarization-diverse CV-QKD system operating over 20 km of standard single-mode fiber using a 10 kHz linewidth free-running local oscillator (LO). To enable secure key generation under such realistic conditions, several classical DSP techniques—originally developed for coherent optical communication systems—were adapted to the quantum regime. Frequency estimation was performed using a multiplexed pilot tone, starting with a coarse peak search in the power spectrum, followed by fine estimation through phase fitting of the filtered pilot. Carrier phase estimation (CPE) was achieved via an Unscented Kalman Filter, enabling robust phase tracking despite the wide laser linewidth. Timing synchronization was implemented using upsampled versions of the transmitted signals. The system also employed state-of-polarization (SOP) compensation using a modified 1-tap CMA, avoiding the need for physical polarization control. In addition, matched filtering using an RRC filter with a 0.2 roll-off factor was applied, and digital whitening and balancing of photodiode responses ensured accurate representation of the vacuum state.  

\begin{figure}[hbt]
    \centering
    \includegraphics[width=0.3\linewidth]{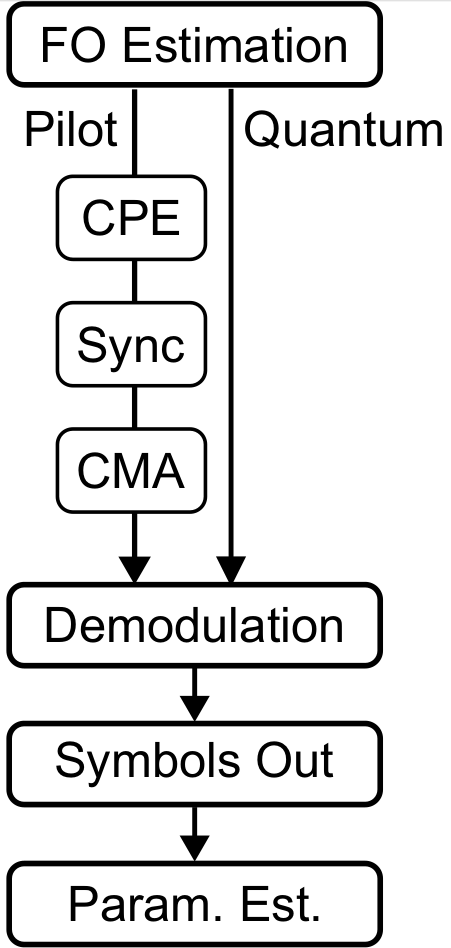}
    \caption{Digital signal processing chain for the CV-QKD system. Reproduced from \cite{chin2024cvqkd}.}
    \label{fig:dsp_chain}
\end{figure}

The complete DSP chain, illustrated in Fig.~\ref{fig:dsp_chain}, clearly summarizes these steps: frequency offset estimation from the pilot, carrier phase equalization via Kalman filtering, SOP compensation with CMA, synchronization, parameter estimation, and final demodulation. This block diagram helps visualize how classical coherent optical DSP modules were re-engineered to operate reliably under quantum-limited conditions.  

These DSP techniques collectively enabled the system to achieve a positive composable secret key rate of $8.7 \times 10^{-3}$ bits per state with $2 \times 10^8$ exchanged quantum states, demonstrating performance comparable to that of narrow-linewidth fiber laser-based setups.  

\begin{figure}[hbt]
    \centering
    \includegraphics[width=1\linewidth]{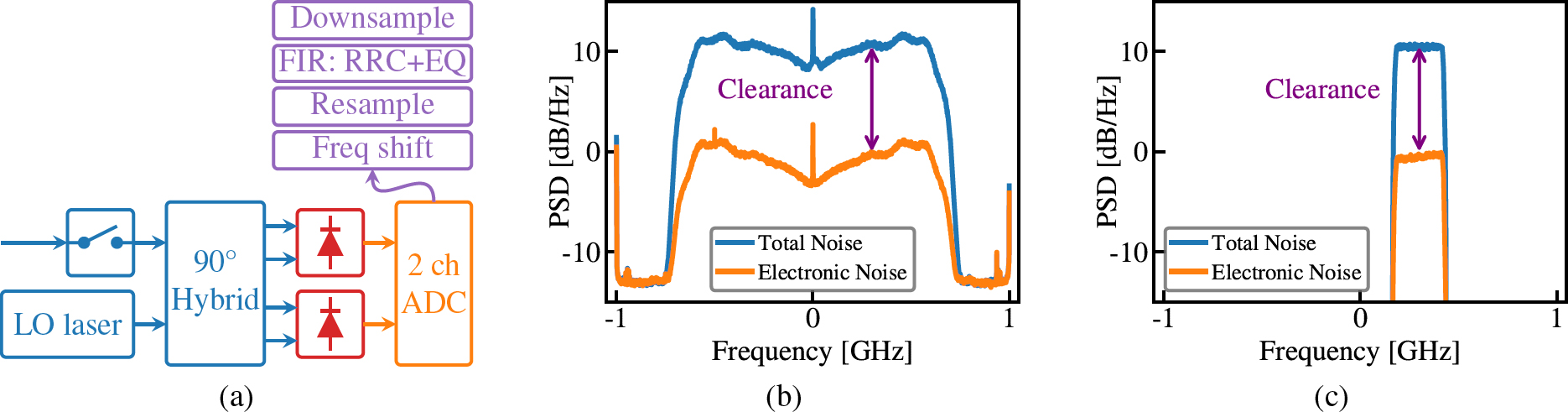}
    \caption{Experimental CV-QKD receiver setup with a conventional coherent detection scheme (a), and measured spectra of total noise before (b) and after (c) DSP processing. Reproduced from \cite{van2023receiver}.}
    \label{fig:receiver_setup}
\end{figure}

In \cite{van2023receiver}, several classical digital signal processing (DSP) techniques, originally developed for coherent optical communication systems, are adapted and applied to enhance the calibration accuracy and temporal stability of CV-QKD receivers. The authors implement frequency shifting to translate the received signal by 300~MHz, thereby mitigating low-frequency interference and enabling cleaner spectral analysis. This step is followed by finite impulse response (FIR) filtering using a root-raised cosine (RRC) pulse shaping filter with 10\% roll-off and static equalization, which together allow for precise bandwidth limitation and improved signal conditioning, even in the absence of a modulated quantum signal. The overall DSP chain is depicted in Fig.~\ref{fig:receiver_setup}, where the experimental CV-QKD receiver architecture and the measured spectra before and after DSP processing are presented. These results illustrate how the combination of frequency translation, filtering, and equalization effectively suppresses spectral artifacts, thereby providing a robust foundation for subsequent excess noise estimation and calibration procedures.

Furthermore, the study introduces the use of the overlapped Allan variance—a technique derived from oscillator frequency stability analysis—as a more statistically robust alternative to conventional pairwise power-difference methods for assessing the temporal stability of receiver noise. Experimental results demonstrate that, while the short-term stability is comparable across distinct local oscillator (LO) lasers, the long-term stability differs by more than an order of magnitude, with high-grade external cavity lasers (ECLs) significantly outperforming standard telecommunication modules. These findings underscore the critical role of DSP-based noise characterization and calibration techniques in ensuring security and optimizing secret key rates in practical CV-QKD implementations.

\begin{figure}[hbt]
    \centering
    \includegraphics[width=1\linewidth]{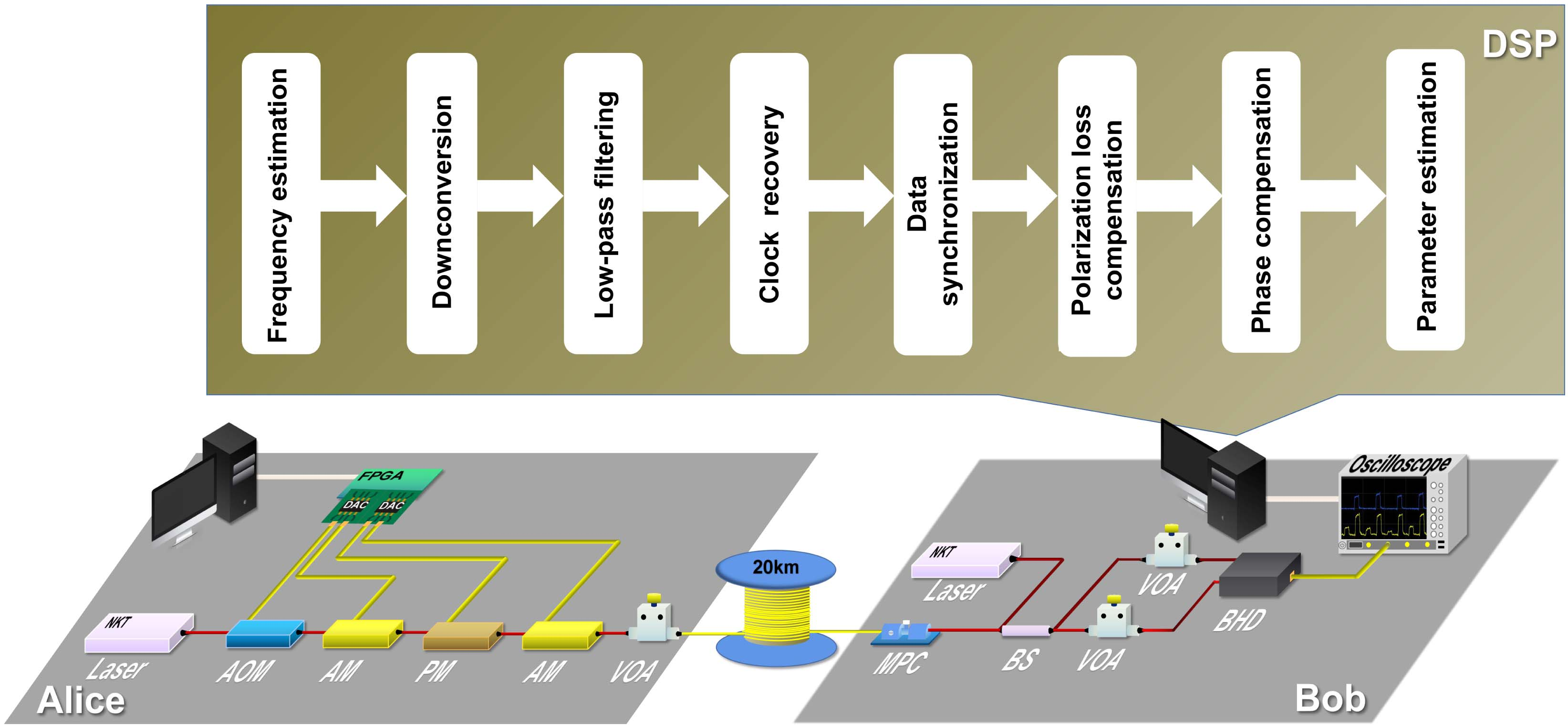}
    \caption{Optical layout and DSP routine of the pilot-sequential LLO GMCS CV-QKD system with heterodyne detection. AOM: acoustic-optic modulator; AM: amplitude modulator; PM: phase modulator; VOA: variable optical attenuator; MPC: manual polarization controller; BS: beam splitter; BHD: balanced homodyne detector; DSP: digital signal processing. Reproduced from \cite{shen2023experimental}.}
    \label{fig:shen_setup}
\end{figure}

In \cite{shen2023experimental}, the authors present an experimental demonstration of a pilot-sequential LLO CV-QKD system based on the GMCS protocol, in which several classical DSP techniques are adapted to quantum communication tasks. The overall optical architecture and digital processing routine are summarized in Fig.~\ref{fig:shen_setup}, where both the transmitter (Alice) and receiver (Bob) subsystems, as well as the subsequent DSP pipeline, are depicted.  

Specifically, a \textit{Kalman filter (KF)} is implemented for \textit{polarization loss compensation (PLC)}, enabling real-time estimation and correction of the polarization-dependent attenuation induced by the manual polarization controller (MPC). This digital compensation considerably improves the stability of the system’s transmittance while maintaining a low excess noise level ($\sim$0.02~SNU) during long-term operation. Moreover, \textit{frequency estimation} via \textit{Fast Fourier Transform (FFT)} is employed to detect the intermediate frequency offset (on the order of 200~MHz) between independent lasers, which is essential for coherent demodulation.  

Subsequently, \textit{orthogonal component extraction} is carried out digitally through \textit{downconversion} and \textit{low-pass filtering}, thereby replacing conventional 90$^\circ$ optical hybrids and eliminating angular measurement errors associated with optical imperfections. For \textit{phase compensation}, a two-stage approach is implemented: \textit{fast-phase estimation} based on pilot pulses, followed by \textit{slow-phase drift correction} using a correlation-based \textit{phase search algorithm}.  

Through the joint application of these DSP techniques, the system demonstrated a stable secret key rate of 110~kbps over a 20~km optical fiber link with 4~dB attenuation, under finite-size block conditions of $10^6$ symbols. This experimental validation highlights the effectiveness of combining simplified optical hardware with advanced DSP methods for practical LLO CV-QKD implementations.  

\subsection{RQ2 – What recent DSP innovations from Coherent Optical Communication have not yet been applied to CV-QKD and could offer significant advances?}

One notable DSP innovation in coherent optical communications that remains unexplored within CV-QKD is the \textit{DSP-Based Physical Layer Security Scheme using OFDM-Driven Phase Modulation}. This technique, as proposed in \cite{he2022dsp}, employs digitally synthesized orthogonal frequency division multiplexing (OFDM) signals as noise-like security keys to drive phase modulators (PMs), enabling both encryption and decryption directly at the physical layer. The scheme introduces dynamic phase and amplitude masking via cascaded dispersive elements (D1, D2) and a PM, offering high entropy and large key spaces without requiring key transmission. When adapted to CV-QKD, this architecture could significantly enhance the security of classical auxiliary channels used for key reconciliation and authentication, which are known to be potential attack vectors in practical QKD systems. Furthermore, the reconfigurable DSP-based key generation may support real-time key management and synchronization in QKD networks, potentially reducing reliance on additional classical infrastructure. The technique's inherent robustness against brute-force and dispersion-compensation attacks---as validated by simulations showing bit error rates (BER) $\geq 0.3$ under unauthorized decryption attempts---suggests that similar physical-layer obfuscation methods could be exploited in CV-QKD to mask calibration phases or mitigate Trojan-horse attacks, thereby increasing overall system resilience. This approach also offers transparency to modulation formats and baud rates, which may facilitate its integration into multiplexed or hybrid classical-quantum transmission systems.

Recent advances in neural network-based nonlinear equalization for coherent optical communication systems, particularly the critical evaluation of machine learning techniques for signal impairment mitigation, present promising opportunities for adaptation in CV-QKD systems. The work~\cite{freire2022neural} provides a comprehensive analysis of deep neural network (DNN) architectures, including multilayer perceptrons (MLP) and bidirectional long short-term memory networks (biLSTM), highlighting key challenges such as overfitting, metric mismatch (e.g., MSE vs. BER), and pitfalls in performance estimation due to periodic data structures and unrealistic assumptions in quality-of-transmission (QoT) metrics. These techniques—collectively described as \textit{neural network-based nonlinear equalization frameworks}—could offer significant improvements in CV-QKD, particularly in the classical post-processing stages. For example, robust equalization using biLSTM networks may enhance quadrature estimation in the presence of nonlinear channel effects and excess noise, improving reconciliation efficiency and potentially increasing the secret key rate. Moreover, their methodology for analyzing the trade-off between equalization performance and computational complexity provides a valuable framework for deploying machine learning components in hardware-constrained CV-QKD receivers. If adapted carefully, such models could mitigate classical channel distortions without violating quantum security assumptions, contributing to more scalable and resilient quantum communication systems.

Another important study is the \textit{Dual-Arms Power Dither Auto Bias Control (ABC)} technique for silicon IQ modulators (SIQMs)~\cite{zhang2022auto}. This method employs differential power dithers applied simultaneously to the upper and lower arms of the I and Q channels, combined with orthogonal integration to precisely lock bias points. Unlike traditional single-arm ABC methods, this approach effectively decouples the bias stability of the phase modulation channel (P) from the thermal and electrical fluctuations of the amplitude modulation channels (I and Q), leading to enhanced stability and reduced bias-induced distortion. In the context of CV-QKD, where IQ modulators are used to prepare coherent quantum states, maintaining highly stable and accurate bias points is essential to preserve state fidelity and minimize excess noise. By ensuring orthogonality and reducing modulation drift—even under temperature fluctuations—this technique could significantly enhance system robustness, improve the precision of Gaussian state preparation, and ultimately increase the achievable secure key rate and distance in CV-QKD implementations. Experimental results in classical systems have shown an Optical Signal-to-Noise Ratio (OSNR) penalty below 0.15~dB at 512~Gb/s DP-16QAM, suggesting that similarly low signal degradation could be expected when adapted to low-power quantum signal generation~\cite{zhang2022auto}.

A promising signal processing technique is the \textit{peak-density K-means clustering-based blind equalization} algorithm~\cite{cao2022optimized}. Originally proposed to address the limitations of conventional blind equalizers such as the Radius-Directed Equalizer (RDE) and Cascaded Multi-Modulus Algorithm (CMMA) in probabilistically shaped high-order QAM signals, this method significantly enhances robustness to amplitude distortion and distribution ambiguity. By identifying the number of amplitude levels and their centroids through local peak-density estimation, followed by a refined K-means clustering step, the algorithm enables more accurate amplitude radius estimation and decision boundaries. In CV-QKD systems---particularly those employing Gaussian-modulated coherent states---the accuracy of parameter estimation (e.g., excess noise variance, channel transmittance) is critical for security and performance. Adapting this technique to CV-QKD could enable more reliable post-processing of quadrature measurements in the presence of detector imperfections, non-Gaussian noise, or amplitude drifts. Such improvements may directly contribute to tighter parameter estimation, enhanced noise tolerance, and ultimately, increased secure key rates in practical CV-QKD implementations.

Another promising study is the Modified Viterbi-Viterbi (MVV) carrier phase recovery algorithm. Originally proposed for high-order modulation formats such as 16-QAM, 64-QAM, and 256-QAM, the MVV algorithm significantly enhances phase noise compensation by utilizing not only the outer constellation points, as in the traditional Viterbi-Viterbi method, but also the intermediate-ring symbols through selective geometric rotation and filtering strategies \cite{zhang2021carrier}. This enables more accurate phase estimation in the presence of laser phase noise, even for linewidths as high as 1 MHz, without substantial degradation in BER. In the context of CV-QKD, where laser phase noise is a critical limiting factor for coherent detection and accurate quadrature estimation, an adapted version of the MVV algorithm could exploit the continuous nature of Gaussian-modulated quadratures to perform robust phase tracking without relying on pilot tones. Such a technique could increase the phase noise tolerance of CV-QKD systems, relax laser linewidth requirements, and thereby extend secure transmission distances and improve secret key rates. 

Another promising recent result is the \textit{Resampling, Retiming, and Equalization} (RRE) algorithm, which integrates these three traditionally separate functions into a single adaptive FIR filter~\cite{song2023low}. Originally developed for low-complexity FPGA implementation in high-speed DP-QPSK systems, the RRE algorithm demonstrated significant reductions in computational complexity---up to 48.3\% compared to time-domain schemes---while improving error vector magnitude (EVM) performance by up to 0.36~dB under fractional oversampling conditions. This unified approach is particularly advantageous in scenarios where sampling rate and hardware resources are constrained. In the context of CV-QKD, the RRE algorithm could offer notable benefits by enhancing temporal synchronization and mitigating ISI, both of which directly impact the precision of quadrature measurements. Its suitability for dispersion-free or short-reach links, such as fiber-based or free-space optical (FSO) CV-QKD systems, aligns well with typical deployment environments. By enabling more accurate sampling and adaptive equalization with minimal hardware overhead, the adoption of the RRE algorithm in CV-QKD receivers could reduce excess noise and enhance secret key rates, thereby improving the overall practicality and efficiency of real-time quantum-secure communication systems.

Recent advances in DSP for coherent optical communications have introduced novel methods based on cyclostationarity theory for chromatic dispersion estimation (CDE) and timing error detection (TED), particularly suited for complex mode-division multiplexing systems. In~\cite{qiao2024novel}, the authors propose the \textit{CAF-4 CDE} and \textit{CAF-4 TED} techniques, which leverage a $4 \times 4$ spectral correlation matrix to provide robust estimation of dispersion and timing errors, even in the presence of severe polarization mode dispersion (PMD) and mode coupling (MC). These methods significantly outperform traditional approaches, achieving chromatic dispersion estimation errors below 50~ps/nm and reducing jitter variance by over 15~dB under MC conditions. While originally developed for orbital angular momentum (OAM) fiber systems, these DSP techniques could also bring significant advantages to CV-QKD. In this context, precise synchronization and dispersion compensation are critical to maintain quadrature measurement integrity and minimizing excess noise. Integrating CAF-4 CDE and TED into CV-QKD receivers may enhance clock recovery and symbol alignment, thereby increasing secret key rates and extending secure transmission distances, particularly in fiber links with non-negligible PMD or CD. This would be especially impactful for future implementations of CV-QKD over multimode or spatially multiplexed networks.

\subsection{RQ3 – What are the future perspectives for using DSP in CV-QKD systems?}

Several recent works provide key insights into how DSP can help overcome limitations in stability, synchronization, and performance under real-world constraints. In~\cite{milovanvcev2021high}, the authors explore DSP techniques for high-rate CV-QKD over mobile WDM fronthaul systems for dense 5G networks, focusing on symbol rates up to 500 MHz using Nyquist pulse shaping and frequency/polarization-multiplexed pilot tones. Their work highlights the importance of precise optical carrier phase tracking and wideband receiver optimization—crucial aspects that require increasingly sophisticated DSP tools to operate within low photon-per-symbol regimes.

~\cite{xu2024robust} presents a robust CV-QKD protocol that achieves high stability even under arbitrary frequency and phase drifts, demonstrating the essential role of DSP for real-time adaptive compensation with complexity independent of block length. The ability to maintain $\eta_q = 1$ (mode matching efficiency) in field scenarios showcases DSP’s potential for long-distance CV-QKD in dynamic environments, especially when moving toward composable security models.

~\cite{chin2022digital} offers an elegant digital synchronization method compatible with local oscillator generation, eliminating the need for costly analog synchronization circuits. Their approach, relying on auxiliary pilot tones without degrading secret key rate (SKR), suggests that future CV-QKD deployments can achieve better cost-efficiency and robustness through DSP-based spectral optimization.

~\cite{wang2023experimental} experimentally demonstrates 5~GBaud CV-QKD using a four-state modulation scheme with DSP-based phase noise and polarization variation compensation. Their system achieves 149.2~Mbps at 10~km, avoiding complex optical setups. This work emphasizes how DSP can replace unstable optical components and paves the way for practical metropolitan QKD deployments with further noise suppression and key rate enhancements.

~\cite{da2024coherent} analyzes modern DSP trends in coherent communications and translates these to CV-QKD, emphasizing that DSP can significantly reduce error correction overhead through well-engineered modulation schemes. Their review underscores the need for field-validation of DSP strategies, especially when implementing discrete modulation protocols and pilot-based synchronization under low SNR conditions.

~\cite{hajomer2024continuous} demonstrate a 10~GBaud CV-QKD system using an integrated photonic-electronic receiver. Their future outlook includes optimizing symbol rate scalability and reducing phase noise via machine learning-based DSP and advanced ADC calibration techniques. Equalization methods such as continuous-time linear equalization (CTLE) and entropic security analysis frameworks are cited as essential future directions.

~\cite{dos2024real} showcases a real-time CV-QKD receiver over FSO urban links, employing GPU-accelerated DSP to maintain SKR stability under atmospheric turbulence. This work illustrates how adaptive DSP algorithms are critical for real-time parameter estimation and noise control, enabling resilient QKD links in harsh and variable environments.

Beyond these detailed contributions, a broad set of studies further supports the transformative role of DSP in CV-QKD. For instance, ~\cite{matsuura2021finite} emphasizes fidelity estimation under finite-size security, while ~\cite{alsalami2022scalar} proposes scalar minimax filtering for phase tracking. ~\cite{tan2024polarization} and ~\cite{pan2023simple} focus on polarization recovery and fast tracking using pilot-assisted methods. Architectures like QOSST~\cite{pietri2024qosst} aim to modularize experimental platforms, while ~\cite{hajomer2022modulation} tackles side-channel mitigation via baseband optical modulation. ~\cite{chin2024composable, chin2024cvqkd} discuss composable security frameworks and low-linewidth laser applicability, and ~\cite{chin2023machine} introduce machine learning for joint polarization and phase estimation. Meanwhile, ~\cite{moreolo2024adoption} and the survey in~\cite{da2024coherent} examine integration of CV-QKD into optical networks and signal processing adaptation to low SNR conditions. Collectively, these works indicate that future DSP development must target scalability, resilience to hardware imperfections, and cross-layer integration—ensuring secure, high-performance CV-QKD systems in heterogeneous communication infrastructures.

\subsection{RQ4 – What performance metrics are most used to evaluate the impact of DSP techniques in CV-QKD?}

A fundamental metric for evaluating the effectiveness of DSP techniques in CV-QKD systems is the \textit{Secret Key Rate} (SKR). This metric quantifies the amount of secret key material that can be generated per unit time under given security assumptions and is commonly used as the primary performance benchmark. Digital signal processing plays a crucial role in optimizing SKR by compensating for transmission impairments such as phase noise, polarization fluctuations, and frequency offsets. Several studies have demonstrated that DSP algorithms—such as carrier recovery, pilot-aided synchronization, and adaptive equalization—are essential to achieving high SKRs across different CV-QKD configurations~\cite{wang2023experimental, shen2023experimental, xu2024robust, aymeric2022quantum}.

Another central performance indicator is the \textit{Excess Noise}, which represents the noise level at the receiver that exceeds the quantum shot noise limit. This metric is tightly linked to the security of the protocol, as it determines the maximum amount of information an eavesdropper could potentially extract. The ability of DSP to suppress or accurately estimate excess noise is therefore critical to system performance. Techniques such as Kalman filtering, digital phase tracking, and optimized filtering structures have been employed to mitigate excess noise arising from hardware imperfections, signal distortion, or classical channel interference~\cite{tan2024polarization, milovanvcev2021high, adillon2024cv, roumestan20226}.

Closely related to these metrics is the \textit{Signal Recovery Accuracy and System Stability}, which assesses the fidelity of the recovered quantum signal and the robustness of the system over time. Effective DSP enables precise reconstruction of transmitted states by compensating for polarization rotation, I/Q imbalance, and carrier phase drift. Experimental implementations relying on training-based equalizers, statistical tests for modulation integrity, and real-time polarization tracking confirm the role of DSP in maintaining the Gaussian nature of the signal and avoiding the need for complex optical compensation techniques~\cite{schiavon2023high, kawakami2024no, pereira2023polarization}.

In multi-channel and integrated environments, \textit{Spectral Efficiency and Channel Coexistence} have become increasingly important performance dimensions. These metrics evaluate the ability of CV-QKD systems to share the optical spectrum with classical channels without degradation. DSP techniques such as pulse shaping, spectral filtering, and cross-talk suppression are key to enabling WDM compatibility and polarization multiplexing. Recent studies have shown that spectral shaping enabled by DSP not only improves efficiency but also facilitates scalable deployment of CV-QKD in existing telecommunication infrastructures~\cite{moreolo2024adoption, iqbal2024sdn, dos2024noise}.

Additionally, auxiliary metrics such as the \textit{Bit Error Rate} (BER), \textit{Quantum Bit Error Rate} (QBER), and \textit{Signal-to-Noise Ratio} are frequently used to evaluate DSP performance at intermediate stages of signal processing. BER is often monitored in classical reference channels to assess the accuracy of phase and frequency estimation, particularly in systems that leverage classical-to-quantum parameter sharing~\cite{aymeric2022quantum}. QBER is relevant in CV-QKD systems employing discrete modulation formats or base reconciliation, serving as an indicator of bit-level fidelity. Meanwhile, the SNR of the quantum signal provides insight into how well the DSP chain preserves quadrature information after channel distortions. Low BER/QBER values and high SNR are indicative of an effective DSP implementation and correlate with improved SKR and protocol robustness~\cite{wang2023experimental, tan2024polarization, kawakami2024no}.

\subsection{RQ5 – What technical challenges still limit the full integration of DSP into CV-QKD systems?}

The full integration of DSP into CV-QKD systems remains hindered by a number of unresolved technical challenges. One of the most persistent limitations lies in the dynamic compensation of polarization drifts and phase noise in real-world fiber channels. The work by ~\cite{chin2024composable} introduces a polarization-diverse heterodyne receiver combined with a modified CMA, enabling single-polarization CV-QKD operation without feedback or calibration, thus maintaining mutual information in the presence of unpredictable polarization rotation. ~\cite{chin2023machine} further explores joint phase and polarization estimation using a machine learning framework based on an UKF, which substantially outperforms CMA in low-SNR regimes, reducing excess noise from $\approx 4.9$ mSNU to $\approx 0.6$ mSNU. Similarly, ~\cite{pereira2023polarization} eliminates the need for real-time feedback and manual alignment through a polarization-diverse receiver architecture capable of passive monitoring and reconstruction of both polarization states. ~\cite{shen2023experimental} complements these approaches with a Kalman filter-based DSP algorithm designed to stabilize secret key rate fluctuations by mitigating residual polarization losses when inexpensive manual controllers are used. ~\cite{pan2023simple} demonstrates a fast polarization tracking algorithm based on orthogonal pilot tones and FIR filters, achieving tracking rates exceeding 12.57 krad/s experimentally, thus addressing rapid polarization perturbations. Additionally, ~\cite{tan2024polarization} develops a time-division multiplexed pilot scheme with ``guard symbols'' to track polarization state variations in GMCS systems without degrading pulse shaping, improving spectral efficiency.

Carrier synchronization and local oscillator (LO) management in ultra-low SNR environments also remain significant bottlenecks. ~\cite{da2024coherent} underscores the importance of robust DSP in compensating for phase/frequency offsets in LLO configurations, highlighting its compatibility with telecom infrastructures. ~\cite{milovanvcev2021high} tackle LLO synchronization by employing dual-tone pilot signals multiplexed in both frequency and polarization, achieving stable operation at 250--500~MHz symbol rates. ~\cite{kawakami2023high} demonstrates long-distance (100~km) CV-QKD by extracting timing information from polarization-multiplexed QPSK pilots, relying on DSP for precise timing recovery. ~\cite{chin2024cvqkd} shows that DSP can handle broader-linewidth commercial ECL lasers (10~kHz) as LOs, thus reducing system cost and complexity. ~\cite{ruckmann2021cv} further reduces the dependence on narrow-linewidth lasers by introducing an extended self-learning Kalman smoother, enabling phase recovery with 300~kHz linewidth lasers in commercial ACO modules. ~\cite{xu2024robust} provides a real-time time-variant DSP scheme with constant $\mathcal{O}(1)$ complexity that stabilizes phase and frequency deviations over long blocks, ensuring low excess noise despite environmental perturbations. ~\cite{aymeric2022quantum} experimentally demonstrates that co-propagating classical data channels can serve as references for LO phase and frequency recovery in CV-QKD, facilitating DSP-based hybrid classical-quantum integration. ~\cite{matalla2023pilot} eliminates the need for pilot tones by implementing a DSP-based timing recovery scheme adapted from Barton and Al-Jalili’s estimator, allowing operation near the shot noise limit even under 10~ppm clock offsets.

On the coexistence front, integrating CV-QKD into dense WDM optical networks poses challenges due to classical channel crosstalk and nonlinear effects. ~\cite{iqbal2024sdn} addresses these with a simulation-based framework that dynamically tunes classical channel parameters via SDN to minimize excess noise in CV-QKD. ~\cite{kawakami2024no} showcases a DSP-enabled CV-QKD system coexisting with 11 WDM channels without guard bands, achieving key rates beyond 75~km. ~\cite{dos2024noise} implement a real-time DSP pipeline (offset correction, pilot-based phase recovery, adaptive filtering) to sustain CV-QKD with 14 adjacent high-power classical channels. ~\cite{adillon2024cv} optimize DSP parameters (e.g., pulse shaping filter roll-off and tap count) to suppress electrical-domain crosstalk in close-frequency multiplexed configurations.

DSP performance is also constrained by the quality of analog components and the level of integration. ~\cite{pietri2023cv} propose a silicon photonic receiver platform with integrated amplification electronics, demonstrating 99.2\% linearity and 1~SNU clearance. ~\cite{milovanvcev2024monolithically} presents a monolithically integrated optoelectronic receiver with a low-noise TIA, reducing input-referenced noise by up to 70\% compared to wire-bonded designs, significantly enhancing the quantum-classical clearance.

Another key limitation lies in modulation formats and post-measurement filtering. ~\cite{roumestan2021high} shows that DSP optimized for probabilistically shaped 64-QAM and 256-QAM formats enables record-high SKRs (over 67~Mbps) over short distances. ~\cite{alsaui2023digital} emphasizes the need for careful post-measurement filtering to preserve shot noise dominance and ensure positive SKRs using commercial hardware. ~\cite{wang2024high} leverages telecom-grade high-bandwidth components in conjunction with a DSP chain tailored for high-rate GMCS CV-QKD. ~\cite{pan2024100} achieves discrete modulation CV-QKD over 100.93~km using probabilistically shaped 16QAM and an advanced DSP pipeline for frequency offset correction, demodulation, and matched filtering. ~\cite{roumestan2024shaped} further explores DSP optimization for PCS-QAM under finite-size effects. To address distance adaptability, ~\cite{berl2024comparison} proposes using DSP-compatible adaptive error-correction codes and modulation variance optimization, demonstrating superior SKRs compared to fixed-rate reconciliation.

Finally, several auxiliary limitations hinder DSP deployment. ~\cite{jin2021key} propose DSP-based and machine learning-assisted sifting strategies (e.g., isolation forest, Wiener filter) to discard anomalous bits and reduce excess noise. ~\cite{van2023receiver} investigates temporal calibration stability through DSP-based Allan variance analysis. ~\cite{pietri2024qosst} introduces QOSST, an open-source DSP platform tailored for CV-QKD at ultra-low SNR, promoting software modularity and hardware abstraction.

\subsection{RQ6 – Which DSP techniques from Coherent Optical Communication have been successfully adapted to CV-QKD, and which required specific modifications?}

In the evolution of CV-QKD, many DSP techniques originating from coherent optical communication have been successfully adapted to the quantum regime. These adaptations often required careful consideration of the physical limitations inherent to quantum systems—particularly the extremely low signal-to-noise ratios (SNR), sensitivity to channel noise, and the need for composable security. Based on the analysis of 31 key publications, these techniques can be grouped into two main categories: (1) directly adapted techniques, and (2) techniques that require specific modifications to function effectively in CV-QKD.

Several DSP methods have been directly adapted from classical systems. Polar coding, a class of capacity-approaching error correction codes introduced by Arıkan, has been employed for information reconciliation in CV-QKD to enhance secret key rates and extend distance performance~\cite{yamaura2024error, mukit2023discrete}. In particular, polar codes were integrated with Cascade protocols to reduce information leakage during error correction.

Another classical technique, Probabilistic Amplitude Shaping (PAS), traditionally used with Quadrature Amplitude Modulation (QAM) to increase spectral efficiency, has been employed to optimize the symbol probability distribution in discrete-modulated CV-QKD~\cite{notarnicola2023probabilistic}. The use of discrete QAM formats combined with PAS allows systems to leverage existing telecommunication infrastructure while improving mutual information.

Multicarrier modulation schemes such as OFDM and OTFS, well-established in high-capacity wireless and fiber-optic links, were shown to be viable for terahertz CV-QKD~\cite{liu2021multicarrier, liu2025otfs}. These schemes improve spectral efficiency and enable robust operation under frequency-selective fading. They required some adjustment, notably in reconciliation algorithms and pilot signal processing, to operate under extreme channel impairments.

Timing synchronization via the Barton and Al-Jalili algorithm has also been adapted to CV-QKD~\cite{matalla2023pilot}. The proposed method facilitates sub-shot-noise clock recovery without the need for pilot tones, extending compatibility with commercial DSP-based coherent receivers.

Several signal estimation and correction methods—such as maximum likelihood estimators (MLEs), Low-Density Parity Check (LDPC) codes, and Toeplitz hashing—have found application in parameter estimation, reconciliation, and privacy amplification stages~\cite{almeida2021secret, mountogiannakis2022composably}. Similarly, phase drift compensation using Particle Filters (PF) and Recursive Least Squares (RLS) algorithms has been demonstrated for systems with independent lasers (LLO CV-QKD)~\cite{qi2024experimental}.

Classical modulation formats such as QPSK were employed in combination with clustering algorithms (e.g., K-means) to post-process coherent measurements in discrete-modulated CV-QKD~\cite{prabhakar2024discrete}. Moreover, machine learning techniques have been repurposed to predict secret key rates in real time with high accuracy, greatly reducing the computational cost compared to numerical simulation~\cite{zhou2022neural}. CUSUM-based security monitoring has also been integrated to detect physical-layer attacks with low latency and high confidence~\cite{gong2021experimental}.

Several other works adapted coherent hardware and classical signal processing to the quantum domain. For instance, phase-sensitive amplifiers (PSAs) were used to increase signal-to-noise ratios, enhancing performance especially in free-space CV-QKD~\cite{alshaer2024enhancing}.

Conversely, a number of DSP techniques required substantial modifications or entirely new frameworks to accommodate the unique requirements of CV-QKD. Rapid polarization tracking, crucial in LLO CV-QKD where state-of-polarization (SOP) drift can induce phase noise, led to the design of a pilot-tone-based tracking algorithm specifically optimized for quantum signals~\cite{pan2023simple}. Similarly, a practical detector model was introduced to account for asymmetric detection imperfections in heterodyne setups~\cite{mi2023continuous}.

Adaptive modulation switching between Gaussian and discrete modulation formats was developed to mitigate phase noise effects and maximize the secret key rate under fluctuating conditions~\cite{alaghbari2021adaptive}. In the binary-modulated CV-QKD regime, a novel Optimal Displacement Threshold Detector (ODTD) was introduced to surpass the conventional 3~dB limit~\cite{zhao2020security, zhao2024security}. This approach was further enhanced with a multi-stage feedforward for improved error performance.

A modified CMA was proposed for SOP estimation in systems using coherent polarization-diverse detection~\cite{chin2024composable}. Similarly, a novel transmitter architecture incorporating frequency- and polarization-multiplexed pilot tones was developed for oscillator synchronization and intradyne coherent detection~\cite{milovanvcev2021high}.

In terahertz CV-QKD, where MIMO architectures are considered, a dedicated channel estimation protocol was proposed to characterize wireless MIMO channels, enabling accurate SVD-based beamforming~\cite{kundu2022channel}. For LLO systems, the Unscented Particle Filter (UPF) algorithm was modified to improve phase noise estimation in non-linear, high-noise environments, outperforming standard Bayesian filters~\cite{ma2025high}.

Efforts were also made to ensure robust detector calibration. A rigorous method to characterize homodyne detection efficiency was introduced to avoid overestimations that could compromise composable security proofs~\cite{zou2022rigorous}. In systems with pilot-signal and quantum-signal polarization multiplexing, a K-means-based demultiplexing method using the Stokes space was proposed for fast and modulation-independent separation~\cite{pan2021polarization}.

Other architectural innovations include time-division dual-quadrature homodyne detection (TDDQHD)~\cite{oh2023continuous}, multidimensional pilot-assisted multiplexing to isolate pilot and quantum channels~\cite{zhang2024novel}, and data-aided time-domain equalization to enhance long-distance performance in low-SNR environments~\cite{pi2023sub}.

To mitigate discretization noise from imperfect modulator resolution, a pre-selection filtering method was introduced before modulation~\cite{wang2022security}. In multimode frequency-multiplexed entanglement sources, cross-talk was compensated using a multimode symplectic transformation tailored to increase mutual information and key rate~\cite{kovalenko2021frequency}. Lastly, a low-complexity frequency-locking algorithm assisted by a pilot tone was designed for heterodyne CV-QKD, integrating DSP-based phase recovery~\cite{ruiz2024low}.

In conclusion, while many DSP techniques from coherent optical communications have been fruitfully transplanted into the CV-QKD framework, a significant subset required fundamental rethinking to align with the security, noise, and hardware constraints of quantum communication.

\section{Conclusions}
\label{conclusions}
This systematic review investigated the application of DSP techniques, originally developed for coherent optical communication systems, to CV-QKD. By employing the APISSER methodology and analyzing 220 peer-reviewed publications from 2021 to 2025, we identified key trends, adaptations, performance impacts, and ongoing challenges associated with this technological convergence.

Our findings indicate that numerous DSP methods—such as phase estimation, polarization tracking, timing synchronization, equalization, and matched filtering—have been successfully adapted to the CV-QKD context, enabling substantial improvements in system performance, particularly in secret key rate, excess noise suppression, and operational stability. Several advanced DSP innovations from coherent optical communication, including machine learning-based equalizers, physical-layer encryption schemes, and unified synchronization-equalization architectures, were found to hold strong potential for future CV-QKD integration, though they remain underexplored in the literature.

The results also reveal that, while many coherent optical communication-derived DSP techniques can be directly applied to CV-QKD with minimal modification, others require substantial redesign to address the stringent noise, security, and hardware constraints inherent in quantum communication. Notably, progress in phase tracking, polarization control, and pilot-assisted synchronization has enabled the use of broader-linewidth lasers and facilitated the transition to fully integrated and scalable receiver architectures.

Despite these advances, several critical challenges remain. These include DSP operation in ultra-low signal-to-noise regimes, compatibility with composable security models, coexistence with classical data channels in WDM networks, and integration with resource-constrained or field-deployable platforms. Addressing these issues will be essential to realizing the full potential of DSP-enhanced CV-QKD systems.

Looking ahead, future research should focus on the co-design of DSP algorithms and quantum system architectures, emphasizing adaptability, modularity, and hardware efficiency. Additionally, the development of standardized DSP frameworks tailored to CV-QKD could accelerate technology transfer and support broader adoption in practical network environments. By building on decades of progress in classical optical communication, DSP offers a promising pathway toward robust, high-rate, and cost-effective quantum-secure communication.

\section*{Acknowledgments}
\label{acknowledgments}
This work was fully funded by the project \textit{HW DSP: Development and Prototyping of Multicore SoC with Dedicated Accelerators and RISC-V DSP}, supported by QuIIN – Quantum Industrial Innovation, the EMBRAPII CIMATEC Competence Center in Quantum Technologies, with financial resources from the PPI IoT/Industry 4.0 of the MCTI, grant number 053/2023, signed with EMBRAPII.

\bibliographystyle{elsarticle-num} 
\bibliography{references}

\end{document}